      [1999/12/01 v1.4c Il Nuovo Cimento]
\documentclass{cimento}

\title{Solar system tests of the cosmological constant}
\author{Philippe Jetzer\from{ins:x} and Mauro Sereno\from{ins:x}}
\instlist{\inst{ins:x} Institut f\"ur Theoretische Physik, Universit\"{a}t Z\"{u}rich,
Winterthurerstrasse 190, 8057 Z\"{u}rich, Switzerland}
\PACSes{\PACSit{04.80.Cc}{95.10.Ce, 95.30.Sf, 96.36.+x}}
\begin{document}

\maketitle

\begin{abstract}
We discuss the influence of the cosmological constant $\Lambda$ on the
gravitational equations of motion of bodies with arbitrary masses and eventually
solve the two-body problem. Observational constraints are derived from
measurements of the periastron advance in stellar systems, in
particular binary pulsars and the solar system. 
For the latter we consider also the change in the mean motion due to $\Lambda$.
Up to now, Earth and Mars data
give the best constraint, $\Lambda~\sim~10^{-36}~\mathrm{km}^{-2}$.
If properly accounting for the
gravito-magnetic effect, this upper limit on $\Lambda$ could greatly
improve in the near future thanks to new data from planned or already
operating space-missions.
Dark matter or modifications of the Newtonian inverse-square law in the solar system 
are discussed as well. Variations in the $1/r^2$ behavior are considered in the 
form of either a possible Yukawa-like interaction or a modification of gravity of MOND type.

\end{abstract}

\section{Introduction}

The understanding of the cosmological constant $\Lambda$ is one of the most outstanding topic in theoretical physics. On the observational side, the cosmological constant is motivated only by large scale structure observations as a possible choice for the dark energy. In fact, when fixed to the very small value of $\sim 10^{-46}~\mathrm{km}^{-2}$, $\Lambda$, together with dark matter, can explain the whole bulk of evidence from cosmological investigations. In principle, the cosmological constant should take part in phenomena on every physical scale but due to its very small size, a local independent detection of its existence is still lacking. Measuring local effects of $\Lambda$ would be a fundamental confirmation and would shed light on its still debated nature, so it is worthwhile to investigate $\Lambda$ at any level. 

Up till now, no convincing method for constraining $\Lambda$ in an Earth's laboratory has been proposed \cite{ref:je+st05}. Astronomical phenomena seem to be more promising. The cosmological constant can affect celestial mechanics and some imprints of $\Lambda$ can influence the motion of massive bodies. In particular, the effect on the perihelion precession of solar system planets has been considered to limit the cosmological constant to $\Lambda \sim 10^{-36}~\mathrm{km}^{-2}$ \cite{ref:je+se06}.

So far local physical consequences of the existence of a
cosmological constant were investigated studying the motion of test
bodies in the gravitational field of a very large mass. This one-body
problem can be properly considered in the framework of the spherically
symmetric Schwarzschild vacuum solution with a cosmological constant,
also known as Schwarzschild-de Sitter or Kottler space-time. The rotation of the central source can also be
accounted for using the so-called Kerr-de Sitter space-time. In \cite{ref:je+se06} we carried out an analysis of the gravitational
$N$-body problem with arbitrary masses in the weak field limit with a
cosmological constant. This study was motivated by the more and more
central role of binary pulsars in testing gravitational and relativistic
effects. 

Gravitational inverse-square law and its relativistic generalization have passed significant tests on very different length- and time-scales. Precision tests from laboratory and from measurements in the solar system and binary pulsars provide a quite impressive body of evidence, considering the extrapolation from the empirical basis \cite{ref:ade+al03,ref:wil06}. First incongruences seem to show up only on galactic scales with the observed discrepancy between the Newtonian dynamical mass and the directly observable luminous mass and they are still in order for even larger gravitational systems. Two obvious explanations have been proposed: either large quantities of unseen dark matter (DM) dominate the dynamics of large systems or gravity is not described by Newtonian theory on every scale.

High precision solar system tests could provide model independent constraints on possible modifications of Newtonian gravity. The solar system is the larger one with very well known mass distribution and can offer tight confirmations of Newtonian gravity and general relativity. Any deviation emerging from classical tests would give unique information either on dark matter and its supposed existence or on the nature of the deviation from the inverse-square law. The orbital motion of solar-system planets has been determined with higher and higher accuracy \cite{ref:pit05b} and recent data allow to put interesting limits on very subtle effects, such as that of a non null cosmological constant 
\cite{ref:je+se06,ref:se+je06b,ref:se+je06c,ref:se+je07}.

\section{The two-body problem}

The total Lagrangian for two particles can be written as
\begin{equation}
{\cal L} \simeq \frac{1}{2}m_\mathrm{a} v_\mathrm{a}^2 +G \frac{m_\mathrm{a} m_\mathrm{b}}{x}+ \frac{1}{2}m_\mathrm{b}
v_\mathrm{b}^2 + \delta {\cal L}_{\mathrm{pN} (\Lambda=0)} + \delta {\cal L}_{\Lambda}
\end{equation}
where $x  \equiv x_\mathrm{a} - x_\mathrm{b} $ is the separation vector and 
$\delta {\cal L}_{\mathrm{pN} (\Lambda=0)}$ and  $\delta {\cal L}_{\Lambda}$ are the pN and $\Lambda$-contributions, respectively, with
\begin{equation}
\delta {\cal L}_{\Lambda} = \frac{\Lambda}{6} (m_\mathrm{a} x_\mathrm{a}^2 + m_\mathrm{b} x_\mathrm{b}^2) .
\end{equation}
Due to cosmological constant, the energy of the system is modified by
a contribution $-\delta {\cal L}_{\Lambda}$. The pN and $\Lambda$
corrections are additive and can be treated separately. 
Since the perturbation due to
$\Lambda$ is radial, the orbital angular momentum is conserved and the
orbit is planar. The main effect of $\Lambda$ on the orbital motion is
a precession of the pericentre.
One gets for the contribution to the precession angular velocity due to $\Lambda$,
\begin{equation}
\dot{\omega}_\Lambda = \frac{\Lambda c^2 P_\mathrm{b}}{4 \pi}
\sqrt{1-e^2},
\end{equation}
where $e$ is the eccentricity and $P_\mathrm{b}$ the Keplerian period of
the unperturbed orbit. This contribution should be considered together with the post-Newtonian
periastron advance, $\dot{\omega}_{\mathrm{pN}} = 3 (2 \pi/P_\mathrm{b})^{5/3} (G
M/c^3)^{2/3} (1-e^2)^{-1} $. The ratio between these two contributions
can be written as,
\begin{equation}
\frac{ \dot{\omega}_\Lambda } {\dot{\omega}_{\mathrm{pN}}} =
\frac{\bar{R}}{R_\mathrm{g}} \frac{\rho_\Lambda }{ \rho } 
= \frac{1}{6}\frac{\bar{R}^4}{R_\mathrm{g}^2} \Lambda  ,
\end{equation}
where $\bar{R} = a (1-e^2)^{3/8}$ is a typical orbital radius with $a$ the semi-major radius of the unperturbed orbit, $\rho \equiv M/(4 \pi \bar{R}^3/3)$ is a typical density of the system and $\rho_\Lambda
\equiv c^2 \Lambda /8 \pi G $ is the energy density associated to
the cosmological constant. The effect of $\Lambda$ can be significant for
very wide systems with a very small mass.

\subsection{Interplanetary measures}

Precessions of the perihelia of the solar system planets have provided
the most sensitive local tests for a cosmological constant so far
\cite{ref:wri98,ref:ker+al03}. Estimates of the anomalous perihelion
advance were recently determined for Mercury, Earth and Mars \cite{ref:pit05a,ref:pit05b}. Such ephemeris were constructed integrating
the equation of motion for all planets, the Sun, the Moon and largest
asteroids and including rotations of the Earth and of the Moon,
perturbations from the solar quadrupole mass moment and asteroid ring
in the ecliptic plane. Extra-corrections to the known general relativistic predictions can
be interpreted in terms of a cosmological constant effect. We considered the 1-$\sigma$ upper bounds. 
Results are listed in Table~\ref{tab:plan}. Best constraints come from Earth and
Mars observations, with $\Lambda \sim 10^{-36}~\mathrm{km}^{-2}$. 
Major sources of systematic errors come from uncertainties about solar oblateness and from the
gravito-magnetic contribution to secular advance of perihelion but
their effect could be in principle accounted for \cite{ref:ior05}. 
The accuracy in determining the planetary orbital
motions will further improve with data from future space-missions. 
By considering a
post-Newtonian dynamics inclusive of gravito-magnetic terms, the
resulting residual extra-precessions should be reduced by several
orders of magnitude, greatly improving the upper bound on $\Lambda$.

\begin{table}
\caption{\label{tab:plan} Limits on the cosmological constant
due to extra-precession of the inner planets of the solar system.}
\begin{tabular}{lrrr}
Name  &  $\delta\dot{\omega}$ (arcsec/year)&
 $\dot{\omega}_\Lambda$ ($\deg$/year)& $\Lambda_\mathrm{lim}~(\mathrm{km}^{-2}) $ \\
\hline
Mercury  &  $-0.36(50)\times 10^{-4}$   &  $9.61{\times} 10^{25} \Lambda /(1~\mathrm{km}^{-2})$ &
$4 {\times} 10^{-35}$
\\
Venus    &  $0.53(30)\times 10^{-2}$     &  $2.51{\times} 10^{26}\Lambda /(1~\mathrm{km}^{-2})$ &
$9 {\times} 10^{-33}$
\\
Earth    &  $-0.2(4)\times 10^{-5}$  & $4.08{\times} 10^{26}\Lambda /(1~\mathrm{km}^{-2})$ &   $1{\times}
10^{-36}$
\\
Mars &      $0.1(5)\times 10^{-5}$  & $7.64{\times} 10^{26} \Lambda /(1~\mathrm{km}^{-2})$& $ 2{\times}
10^{-36}$
\\
\end{tabular}

\end{table}

\subsection{Binary pulsars}

Binary pulsars have been providing unique possibilities of probing
gravitational theories. The advance of periastron of
the orbit, $\dot{\omega}$, depends on the total mass of the system and
on the cosmological constant. In principle, because Keplerian orbital parameters
such as the eccentricity $e$ and the orbital period $P_\mathrm{b}$ can
be separately measured, the measurement of $\dot{\omega}$ together
with any two other post-Keplerian parameters would provide three
constraints on the two unknown masses and on the cosmological
constant. As a matter of fact for real systems, the effect of $\Lambda$ is much smaller
than $\dot{\omega}_{\mathrm{pN}}$, so that only upper bounds on the
cosmological constant can be obtained by considering the uncertainty
on the observed periastron shift \cite{ref:je+se06}. Despite of the low accuracy in the
measurement of $\dot{\omega}$, PSR J1713+0747 provides the best
constraint on the cosmological constant, $\Lambda \sim 8
{\times}10^{-30}~\mathrm{km}^{-2}$. Uncertainties as low as $\delta\dot\omega
\geq 10^{-6}$ have been achieved for very well observed systems, such
as B1913+16 and B1534+12. Such an accuracy for B1820-11 would allow
to push the bound on $\Lambda$ down to $10^{-33}~\mathrm{km}^{-2}$.

Better constraints could be obtained by determining post-Keplerian
parameters in very wide binary pulsars. We examined systems with known
period and eccentricity. The binary pulsar
having the most favorable orbital properties for better constraining
$\Lambda$ is the low eccentricity B0820+02, located in the Galactic
disk, with $\dot{\omega}_\Lambda
\sim 1.4{\times}10^{27}\Lambda / (1~\mathrm{km}^{-2})\deg/\mathrm{days}$. 
For binary pulsars J0407+1607, B1259-63, J1638-4715 and J2016+1948,
the advance of periastron due to the cosmological constant is between
7 and $9{\times}10^{26}\Lambda/ (1~\mathrm{km}^{-2}) \deg/\mathrm{days}$. All of these shifts are
of similar value or better than the Mars one. A determination of
$\dot{\omega}$ for B0820+02 with the accuracy obtained for B1913+16, i.e.
$\delta \dot{\omega}\geq ~10^{-6} \deg/\mathrm{days}$  would allow to
push the upper bound down to $ 10^{-34}-10^{-33}~\mathrm{km}^{-2}$.

\section{Mean motion}

\begin{table}
\caption{\label{tab:meanN} Limits on the cosmological constant
due to anomalous mean motion of the solar system planets. $\delta a$ is the statistical
error in the orbital semimajor axis; $\Lambda_\mathrm{lim}$ is the 
$1-\sigma$ upper bound on the cosmological constant.}
%\begin{ruledtabular}
\begin{tabular}{lrr}
Name  &  $ \delta a~(\mathrm{km})$ & $\Lambda_\mathrm{lim}~(\mathrm{km}^{-2})$ \\
\hline
Mercury  &  $0.105 \times 10^{-3}$   &  $ 1 {\times} 10^{-34}$
\\
Venus    &  $0.329 \times 10^{-3}$   &  $ 3 {\times} 10^{-35}$
\\
Earth    &  $0.146 \times 10^{-3}$   &  $ 4 {\times} 10^{-36}$
\\
Mars &      $0.657 \times 10^{-3}$   &  $ 3 {\times} 10^{-36}$
\\
Jupiter &   $0.639  \times 10^{+0}$   &  $ 2 {\times} 10^{-35}$
\\
Saturn  &   $0.4222 \times 10^{+1}$   &  $ 1 {\times} 10^{-35}$
\\
Uranus  &   $0.38484\times 10^{+2}$   &  $ 8 {\times} 10^{-36}$
\\
Neptune &   $0.478532\times 10^{+3}$   &  $ 2 {\times} 10^{-35}$
\\
Pluto   &  $0.3463309\times 10^{+4}$  &  $ 4 {\times} 10^{-35}$
\end{tabular}
%\end{ruledtabular}

\end{table}

A positive cosmological constant would decrease the effective mass of the Sun as seen by the outer planets. Due to $\Lambda$, 
the radial motion of a test body around a central mass $M$ is affected by an additional acceleration, $a_{\Lambda}
=\Lambda c^2 r/3$,  and a change in the Kepler's third law occurs \cite{ref:wri98}. For a circular orbit,
\begin{eqnarray}
\omega^2 r & =&  \frac{G M}{r^2} - \frac{\Lambda c^2}{3} r  \label{mean1} \\
& \equiv & \frac{G M_\mathrm{eff}}{r^2} . \label{mean2}
\end{eqnarray} where $\omega$ is the angular frequency. By comparing Eqs.~(\ref{mean1},~\ref{mean2}), 
we get the variation due to $\Lambda$ in the effective mass for test bodies at radius $r$,
\begin{equation}
\label{mean3}
\frac{\delta M_\mathrm{eff}}{M} = - \frac{1}{3}\Lambda \frac{r^3}{r_\mathrm{g}},
\end{equation}
where $r_\mathrm{g}=GM/c^2$ with $M$ the value of the central mass.
In other words, the mean motion $n \equiv \sqrt{G M/ a^3}$ is changed by \cite{ref:wri98},
\begin{equation}
\label{mean4}
\frac{\delta n}{n} = - \frac{\Lambda}{6} \frac{a^3}{r_\mathrm{g}}.
\end{equation}

We can then evaluate the statistical error on the mean motion for each major planet, $\delta n = - (3/2) n \delta a/a$, and translate it into an uncertainty on the cosmological constant. Results and are listed in Table~\ref{tab:meanN}. Best limits comes from Earth and Mars. Errors in Table~\ref{tab:meanN} are formal and could be underestimated. Current accuracy can be determined evaluating the discrepancies in different ephemeris \cite{ref:pit05b}. Differences in the heliocentric distances do not exceed $10~\mathrm{km}$ for Jupiter and amount to 180, 410, 1200 and 14000~km for Saturn, Uranus, Neptune and Pluto, respectively \cite{ref:pit05b}. Bounds on $\Lambda$ from outer planets reported in Table~\ref{tab:meanN} should be accordingly increased.

Unlike inner planets, radiotechnical observations of outer planet are still missing and their orbits can not 
be determined with great accuracy. We can assume a conservative uncertainty of $\delta a \sim 10^{-1}$-$1~\mathrm{km}$ on the 
Neptune or Pluto orbits from future space missions, which would bound the cosmological constant 
to $\Lambda \leq 10^{-38}$-$10^{-39}~\mathrm{km}^{-2}$, three order of magnitude better than today's constraints from Mars.

Pioneer spacecrafts have been considered as ideal systems to perform precision celestial mechanics experiments \cite{ref:and+al02}. Analyzed data cover a heliocentric distance out to $\sim 70~\mathrm{AU}$ and show an anomalous acceleration directed towards the sun with a magnitude of $\sim 9 {\times} 10^{-8}~\mathrm{cm~s}^{-2}$ \cite{ref:and+al02}. If all the systematics were accounted for, that acceleration could be originated by some new physics. An interpretation of these data in terms of $\Lambda$ would imply a negative cosmological constant, which seems quite unlikely. Taken at the face value, the Pioneer anomalous acceleration would give $\Lambda \sim -3 \times 10^{-35}~\mathrm{km}^{-2}$.

\section{Dark matter}

In the dark matter scenario the Milky Way is supposed to be embedded in a massive dark halo, with the 
local DM density at the solar circle 
of about $\rho_\mathrm{DM} \sim 0.2 \times 10^{-21}~\mathrm{kg/m}^3$, in excess of nearly five orders of magnitude with 
respect to the mean cosmological dark matter density.

Galactic dark matter can cause extra-perihelion precession in the solar system, which assuming a constant density  
(locally in the solar system) $\rho_\mathrm{DM}$, induces a perturbing radial acceleration 
$\delta {\cal A}_\mathrm{R} = -(4\pi G \rho_\mathrm{DM}/3) r$ at radius $r$. This leads 
(by also averaging over a period) to an extra-precession rate that can be written as

\begin{equation}
\langle \dot{\omega}_\mathrm{p} \rangle = - \frac{2 G \pi \rho_\mathrm{DM}}{n}\left(1-e^2\right)^{1/2}.
\end{equation}

Note that for an effective uniform density of matter represented by a cosmological constant, i.e. $ \rho_\mathrm{DM} =-c^2 \Lambda/(4\pi G)$, the classical result for orbital precession due to $\Lambda$ is retrieved \cite{ref:ker+al03}. 
The best upper bound on local dark matter density comes from Mars data, see Table~\ref{tab:peri}. 
The accuracy on Mars precession should improve by more than six orders of magnitude to get constraints competitive with local estimates based on Galactic observables.

Bounds on $\rho_\mathrm{DM}$ from deviations in the mean motion of inner planets, see Table~\ref{tab:mean}, are of the same order of magnitude of constraints from extra-precession. Observations of outer planets provide constraints that are an order of magnitude larger but they give the best future prospects. Since the required accuracy to probe the effects of a given uniform background decreases as $\propto a^{-4}$, whereas the measurements precision of ranging observations is roughly proportional to the range distance, exploration of outer planets seems pretty interesting. Dark matter with $\rho_\mathrm{DM} \simeq 0.2 \times 10^{-21}~\mathrm{kg/m}^3$ could be detected if the orbital axis of the Uranus, Neptune and Pluto orbits were determined with an accuracy of $\delta a \sim 3 \times 10^{-2},~ 2\times 10^{-1}$ and $5\times 10^{-1}$~m, respectively. 
Up till now, the only ranging measurements available for Uranus and Neptune are the Voyager 2 flyby data, 
with an accuracy in the determination of distance of $\sim 1~\mathrm{km}$ \cite{ref:and+al95}, not so far 
from what required to probe solar system effects of dark matter.

\begin{table}
\caption{\label{tab:peri} 2-$\sigma$ constraints from extra-precession of the inner planets of the solar system. $\delta\dot{\omega}_\mathrm{p}$ is the observed extra-precession rate; $\delta {\cal A}_\mathrm{R}$  is a constant perturbative radial acceleration at the planet orbit and $\rho_\mathrm{DM}$ is the DM density within the planet orbit.}
\begin{tabular}{lrcl}
\hline
Name  &  $\delta\dot{\omega}_\mathrm{p}$~(arcsec/year) &
 $ \delta {\cal A}_\mathrm{R}~ (\mathrm{m}/\mathrm{s}^2) $  &  $ \rho_\mathrm{DM}~(\mathrm{kg}/\mathrm{m}^3)$ \\
\hline
Mercury  &  $-0.36(50)\times 10^{-4}$   &  $-1 {\times} 10^{-12} \leq \delta {\cal A}_\mathrm{R} \leq 5 {\times} 10^{-13}$   &   $ < 4 {\times} 10^{-14}$
\\
Venus    &  $0.53(30)\times 10^{-2}$    &  $-4 {\times} 10^{-12} \leq \delta {\cal A}_\mathrm{R} \leq 6 {\times} 10^{-11}$   & $ < 8 {\times} 10^{-14}$
\\
Earth    &  $-0.2(4)\times 10^{-5}$     &  $-5 {\times} 10^{-14} \leq \delta {\cal A}_\mathrm{R} \leq 3 {\times} 10^{-14}$  & $ <7{\times} 10^{-16}$
\\
Mars &      $0.1(5)\times 10^{-5}$      &  $-3 {\times} 10^{-14} \leq \delta {\cal A}_\mathrm{R} \leq 4 {\times} 10^{-14}$  &  $ < 3 {\times} 10^{-16}$
\\
\hline
\end{tabular}
\end{table}

\begin{table}
\caption{\label{tab:mean} 2-$\sigma$ upper bounds from anomalous mean motion of the solar system planets; 
$\delta a$ is the uncertainty on the semimajor axis; $\delta {\cal A}_\mathrm{R}$ is an anomalous 
constant radial acceleration; $\rho_\mathrm{DM}$ is the dark matter density.}
\begin{tabular}{lrrl}
\hline
Name  &  $ \delta a~(\mathrm{m})$  & $ |\delta {\cal A}_\mathrm{R} |~ (\mathrm{m}/\mathrm{s}^2) $  &  
$ \rho_\mathrm{DM}~(\mathrm{kg}/\mathrm{m}^3)$ \\
\hline
Mercury  &  $0.105 \times 10^{+0}$   &  $ \leq 4 {\times} 10^{-13}$   &   $  \leq 3 {\times} 10^{-14}$ \\
Venus    &  $0.329 \times 10^{+0}$   &  $ \leq 2 {\times} 10^{-13}$   &   $  \leq 7 {\times} 10^{-15}$ \\
Earth    &  $0.146 \times 10^{+0}$   &  $ \leq 3 {\times} 10^{-14}$   &   $  \leq 8 {\times} 10^{-16}$  \\
Mars &      $0.657 \times 10^{+0}$   &  $ \leq 4 {\times} 10^{-14}$   &   $  \leq 7 {\times} 10^{-16}$ \\
Jupiter &   $0.639  \times 10^{+3}$  &  $ \leq 1 {\times} 10^{-12}$   &   $  \leq 5 {\times} 10^{-15}$\\
Saturn  &   $0.4222 \times 10^{+4}$  &  $ \leq 1 {\times} 10^{-12}$   &   $  \leq 3 {\times} 10^{-15}$\\
Uranus  &   $0.38484\times 10^{+5}$  &  $ \leq 1 {\times} 10^{-12}$   &   $  \leq 2 {\times} 10^{-15}$ \\
Neptune &   $0.478532\times 10^{+6}$ &  $ \leq 4 {\times} 10^{-12}$   &   $  \leq 3 {\times} 10^{-15}$  \\
Pluto   &  $0.3463309\times 10^{+7}$ &  $ \leq 1 {\times} 10^{-11}$   &   $  \leq 8 {\times} 10^{-15}$  \\
\hline
\end{tabular}
\end{table}

\section{MOND}

MOND theory was initially proposed either as a modification of inertia or of gravity \cite{ref:mil83}. 
According to this second approach, the gravitational acceleration $\mathbf{g}$ if related to the Newtonian 
gravitational acceleration $\mathbf{g}_\mathrm{N}$ as
\begin{equation}
\label{mond1}
 \mu (|\mathbf{g}|/a_0)\mathbf{g} = \mathbf{g}_\mathrm{N}
\end{equation}
where $a_0$ is a physical parameter with units of acceleration and $\mu (x)$ is an unspecified function which runs from $\mu (x)=x $ at $x \ll 1$ to $\mu (x) = 1 $ at $x \gg 1$. Whereas the Newtonian trend is recovered at large accelerations, in the low acceleration regime the effective gravitational acceleration becomes $g \simeq \sqrt{g_\mathrm{N} a_o}$. The asymptotically flat rotation curves of spiral galaxies and the Tully-Fisher law are explained by such a modification and a wide range of observations is fitted with the same value of $a_0 \simeq 1.2 \times 10^{-10}~\mathrm{m~s}^{-2}$ \cite{ref:sa+mc02}. 

The $\mu$ function is formally free but, as a matter of facts, fits to rotation curves or considerations on the external field effects suggest a fairly narrow range. The standard interpolating function proposed by \cite{ref:mil83},

\begin{equation}
\label{mond2}
\mu (x) = x/\sqrt{1+x^2} ,
\end{equation}

provides a reasonable fit to rotation curves of a wide range of galaxies. Based on a detailed study of the velocity curves of the Milky Way and galaxy NGC~3198, it has also been proposed the alternative interpolating function,
\begin{equation}
\label{mond3}
\mu (x) = x/(1+x) .
\end{equation}

For a quite general class of interpolating functions, we can write \cite{ref:mil83}
\begin{equation}
\label{mond4}
\mu (x) \simeq 1-k_0 (1/x)^m ,
\end{equation}
which leads to the modified gravitational field \cite{ref:tal+al88}

\begin{equation}
\label{mond5}
\mathbf{g} \simeq \mathbf{g}_\mathrm{N}\left[ 1+ k_0 (a_0/ |\mathbf{g}_\mathrm{N}|)^m \right] .
\end{equation}
For $x \gg 1$, Eq.~(\ref{mond2}) and Eq.~(\ref{mond3}) can be recovered for $\left\{ k_0, m\right\}=\{1/2,2 \}$ and $\{1 ,1\}$, respectively.

The rate of perihelion shift in the Newtonian regime of MOND ($x \gg 1$) with a generic interpolating function in the form of Eq.~(\ref{mond4}) can be expressed in terms of hyper geometric functions, which for a small eccentricity is given by
\begin{eqnarray}
\label{peri4}
\langle \dot{\omega}_\mathrm{p} \rangle & = & - k_0 n \left( \frac{a}{r_\mathrm{M}}\right )^{2 m} m \nonumber \\ 
& \times & \left\{ 1+ e^2 [ 1-m(5-2m)]/4 + {\cal O}(e^4) \right\},
\end{eqnarray}
where $r_\mathrm{M} \equiv \sqrt{G M/a_0}$. As for the DM case, the Mars data is the more effective in constraining 
the parameter space \cite{ref:se+je06c}. For $k_0 \sim 1$, we get $m \geq 1.5$. 

Results from analysis of mean motion are similar to extra-precession analysis. The interpolating function in Eq.~(\ref{mond3}) is not consistent with solar system data. From Uranus data, we get  $m \geq 1.4$ assuming $k_0 \simeq 1$. Again, best future prospects are related to radio-technical determination of orbits of outer planets. The standard interpolating function in Eq.~(\ref{mond2}) could be (dis-)probed if the axes of the Uranus, Neptune and Pluto orbit were determined with an accuracy of $\delta a \sim 3 \times 10,~ 3\times 10^{2}$ and $ 1\times 10^{3}$~m, respectively.

\section{Yukawa-like fifth force}

We consider an additional contributions to the gravitational potential in the form of a Yukawa-like term. 
The weak field limit of the gravitational potential, $\phi$, can be written as a sum of a Newtonian 
and a Yukawa-like potential, for a point mass $M$,
\begin{equation}
\label{ykw1}
\phi  = -\frac{G_\infty M}{r}\left[ 1+ \alpha_\mathrm{Y}  \exp \left\{ -\frac{r}{\lambda_\mathrm{Y}}\right\}\right],
\end{equation}
where $\alpha_\mathrm{Y}$ is a dimensionless strength parameter and $\lambda_\mathrm{Y}$ is a length cutoff. 

The potential in Eq.~(\ref{ykw1}) goes as $\propto 1/r$ both on a small scale ($ r \ll \lambda_\mathrm{Y}$), with an effective coupling constant $G_\infty (1 +\alpha_\mathrm{Y})$, and on a very large scale, where the effective gravitational constant is $G_\infty $. We will take $G_\infty = G /(1+\alpha_\mathrm{Y})$, so that the value of the coupling constant on a very small scale matches the observed laboratory value, $G$. The total gravitational acceleration felt by a planet embedded in the potential~(\ref{ykw1}) can be written as,
\begin{equation}
\mathbf{g} = -\hat{\mathbf{r}}\frac{G_\infty M}{r^2}\left[ 1+\alpha_\mathrm{Y} \left(1+\frac{r}{\lambda_\mathrm{Y}} \right)  \exp \left\{ -\frac{r}{\lambda_\mathrm{Y}}\right\}  \right] .
\end{equation}
For $\alpha_\mathrm{Y}<0(>0)$, gravity is enhanced (suppressed) on a large scale.

The anomalous precession rate due to a Yukawa-like contribution to the gravitational potential is
\begin{eqnarray}
\label{peri5}
\langle \dot{\omega}_\mathrm{p} \rangle & = & \alpha_\mathrm{Y} \left( \frac{a}{\lambda_\mathrm{Y}} \right)^2 \exp\left\{ -\frac{a}{\lambda_\mathrm{Y}}\right\}\frac{n}{2} \\
& \times & \left\{ 1-\frac{1}{8} \left[ 4 - \left(\frac{a}{\lambda} \right)^2 \right]e^2  + {\cal O} (e^4)\right\}. \nonumber
\end{eqnarray}

Extra-precession data for a planet with semimajor axis $a$ mainly probe scale lengths of 
$\lambda_\mathrm{Y} \sim a/2$. Solar system data allow to constrain departures from the inverse-square 
law with high accuracy for a scale length $\lambda_\mathrm{Y} \sim 10^{10}-10^{11}$~m 
\cite{ref:tal+al88,ref:ior05}. Bounds are mainly determined from Mercury and Earth data. 
For $\lambda_\mathrm{Y} \sim 10^{11}$~m, we get $-5 \times 10^{-11} \leq \alpha_\mathrm{Y} \leq 6 \times 10^{-11}$.

Comparison of Keplerian mean motions of inner and outer planets can probe a Yukawa-like contribution only if planets feel 
different effective gravitational constants. Such test is insensitive to values of $\lambda_\mathrm{Y}$ either much 
less the orbit radius of the inner planet or much larger than the orbit of the outer planets \cite{ref:ade+al03}. 
Differently from extra-precession of perihelion, which appears only for departures from the inverse square law, changes 
in the mean motion can 
appear even if both planets feel a gravitational acceleration $\propto 1/r^2$  but with different 
renormalized gravitational constants. Considering inner planets, Earth data give 
$|\alpha_\mathrm{Y}| \leq 6 \times 10^{-12}$ for $\lambda_\mathrm{Y} \leq 2 \times 10^{10}$~m. The best constraint from outer 
planets is due to Jupiter, with $|\alpha_\mathrm{Y}| \leq 5 \times 10^{-9}$ for $\lambda_\mathrm{Y} \leq 10^{11}$~m.

\section{Conclusions}

We determined observational limits on the cosmological constant
from measured periastron shifts. With respect to previous
similar analyses performed in the past on solar system planets,
our estimate was based on a recent determination of the planetary
ephemeris properly accounting for the quadrupole moment of the Sun
and for major asteroids. The best constraint comes from Mars and Earth,
$\Lambda \leq 1-2\times 10^{-36}~\mathrm{km}^{-2}$.
Due to the experimental accuracy, observational limits on
$\Lambda$ from binary pulsars are still not competitive with
results from interplanetary measurements in the solar system. 
The bound on $\Lambda$ from Earth or Mars perihelion shift is nearly $\sim
10^{10}$ times weaker than the determination from observational cosmology, $\Lambda
\sim 10^{-46}~\mathrm{km}^{-2}$, but it still gets some relevance. The
cosmological constant might be the non perturbative trace of some
quantum gravity aspect in the low energy limit \cite{ref:pad05}. $\Lambda$
is usually related to the vacuum energy density, whose properties
depends on the scale at which it is probed \cite{ref:pad05}. So that, in
our opinion, it is still interesting to probe $\Lambda$ on a scale of
order of astronomical unit. Measurements of periastron shift should be
much better in the next years. New data from space-missions should get
a very high accuracy and might probe spin effects on the orbital
motion \cite{ref:oco04,ref:ior05}. A proper consideration of the
gravito-magnetic effect in these analyses plays a central role to
improve the limit on $\Lambda$ by several orders of magnitude.

Debate between dark matter and departures from inverse-square law is still open. Results are still non-conclusive but nevertheless interesting. Best constraints come from perihelion precession of Earth and Mars, with similar results from modifications of the third Kepler's law. The upper bound on the local dark matter density, $\rho_\mathrm{DM} \leq 3 {\times} 10^{-16}~\mathrm{kg/m}^3$, falls short to estimates from Galactic dynamics by six orders of magnitude.  

Deviations of the gravitational acceleration from $1/r^2$ are really negligible in the inner regions. A Yukawa-like fifth force is strongly constrained on the scale of $\sim 1$~AU. For a scale-length $\lambda_\mathrm{Y} \sim 10^{11}~\mathrm{m}$, a Yukawa-like modification can contribute to the total gravitational action for less then one part on $10^{11}$.

A large class of MOND interpolating function is excluded by data in the regime of strong gravity. The onset of the asymptotic $1/r$ acceleration should occur quite sharply at the edge of the solar system, excluding the more gradually varying $\mu(x)$ suggested by fits of rotation curves. On the other hand, the standard MOND interpolating function $\mu(x) = x/(1+x^2)^{1/2}$ is still in place. Studies on planetary orbits could be complemented with independent observations in the solar system. Mild or even strong MOND behavior might become evident near saddle points of the total gravitational potential, where MONDian phenomena might be put at the reach of measurements by spacecraft equipped with sensitive accelerometers. 
As a matter of fact, fits to galactic rotation curves, theoretical considerations on the external field 
effects and solar system data could determine the shape of the interpolating 
function with a good accuracy on a pretty large intermediate range between the deep Newtonian and MONDian asymptotic behaviors.

Future experiments performing radio ranging observations of outer planets could greatly improve our knowledge about gravity in the regime of large accelerations. The presence of dark matter could be detected with a viable accuracy of few tenths of a meter on the measurements of the orbits of Neptune or Pluto, whereas an uncertainty as large as hundreds of meters would be enough to disprove some pretty popular MOND interpolating functions. 
                           
\acknowledgments                          
PhJ would like to thank the organizers of the I Italian-Pakistan Workshop
on Relativistic Astrophysics for the invitation and for the nice
atmosphere of the meeting.

\newpage

\end{document}